# Electro-thermal simulation of superconducting nanowire avalanche photodetectors

Francesco Marsili[1,*], Faraz Najafi[1], Charles Herder[1], Karl K. Berggren[1,2]

[1]*Department of Electrical Engineering and Computer Science, Massachusetts Institute of Technology, 77 Massachusetts Avenue, Cambridge, Massachusetts 02139, USA.*

[2]*Kavli Institute of Nanoscience, Delft University of Technology, Lorentzweg 1, 2628CJ Delft, The Netherlands.*

Equation Chapter 1 Section 1

We developed an electro-thermal model of NbN superconducting nanowire avalanche photodetectors (SNAPs) on sapphire substrates. SNAPs are single-photon detectors consisting of the parallel connection of $N$ superconducting nanowires. We extrapolated the physical constants of the model from experimental data and we simulated the time evolution of the device resistance, temperature and current by solving two coupled electrical and thermal differential equations describing the nanowires. The predictions of the model were in good quantitative agreement with the experimental results.

We simulated the photoresponse of NbN superconducting nanowire avalanche photodetectors (SNAPs) [1] on sapphire substrates. SNAPs are single-photon detectors consisting of the parallel connection of $N$ superconducting nanowires ($N$-SNAPs, see Figure 1), which provide a signal-to-noise ratio (*SNR*) factor $\sim N$ higher than ordinary superconducting nanowire single-photon detectors (SNSPDs) [1], consisting of a single nanowire. Our group recently demonstrated that this enhancement of the *SNR* was crucial to reading out the photoresponse of ultra-narrow (30- and 20-nm-wide) nanowires [2].

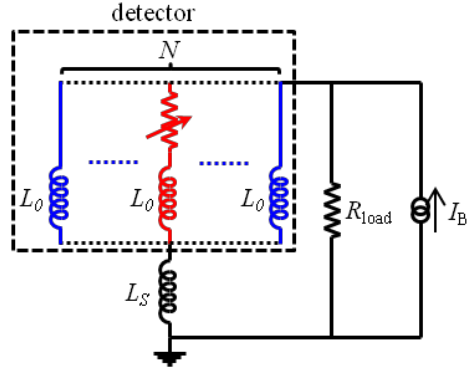

Figure 1. Equivalent electrical circuit of an $N$-SNAP. The initiating section is in red, the secondary sections are in blue. $R_{load}$ = 50 Ω, which models the input impedance of the RF amplifier used to read-out the detector signal [3].

The detector is connected in series with an inductor ($L_S$), to insure the correct operation of the device and in parallel with a readout resistor $R_{load}$ (Figure 1). As all the parallel sections are nominally equal and have the same kinetic inductance ($L_0$, see Figure 1), at the steady state they are biased at the same current ($I_B / N$, where $I_B$ is the total bias current through the device). In this work, $L_S$ was designed to satisfy the condition $L_S (N - 1) / L_0 = 10$. A study of the relation between the value of $L_S$ and the device performance will be reported elsewhere.

When one section absorbs a photon, it switches to the normal state (initiating section) and diverts part of its current to the remaining sections (secondary sections) and $R_{load}$. If $I_B$ is higher than a particular current level (the avalanche threshold current, $I_{AV}$), the current redistributed from the initiating section biases the secondary sections above their critical current $I_C$. At this point, all the sections become resistive, so most of the current flowing through the device $I_B$ is diverted to $R_{load}$, where a current pulse $\sim N$ times higher than the current in a single section is formed.

$I_{AV}$ is a key parameter for the device design and can be easily measured experimentally, providing a convenient validity check for the model of the device physics. As the existing model of SNAP operation failed to fit the experimental data [2, 4], we developed an alternative phenomenological model describing the dynamics of the circuit after the formation of a normal domain in one of the sections of the device.

We solved the two coupled electrical and thermal differential equations reported in ref. [5] for each section of the SNAP at every instant of the simulation, obtaining the time evolution of the currents in the circuit and of the nanowire resistances and temperature profiles. Prior models for superconducting-nanowire-based detectors were not suitable to accurately describe the cascade switching of the secondary sections in SNAPs because they either (1) disregarded the expansion of the photon-induced normal domain due to Joule heating [6]; (2) could only model single-nanowire detectors [5, 7]; (3) described parallel-nanowire detectors with a purely-electrical model [8]; or (4) described SNAPs as equivalent wider-nanowire single-nanowire detectors [9].

The main assumptions we made to model the photoresponse of SNAPs are discussed in the following paragraphs (see supplementary information).

Following ref. [5], we assumed the thermal response of the NbN nanowire to be bolometric and described the electron and phonon sub-systems with a single reduced temperature. We did not use a two-temperature (2-$T$) description of our system such as the one proposed in ref. [7] for the following reasons. First, the phonon escape time of few-nm-thick NbN on sapphire ($\tau_{es}$) is of the same order than the phonon-electron interaction time ($\tau_{p-e}$) [10], which makes assuming thermal equilibrium between the two sub-systems a reasonable approximation (which does not apply e.g. to Nb, for which $\tau_{p-e} \sim 10\ \tau_{es}$ [10]). Second, the 2-$T$ model relies on six parameters which are extremely challenging to determine experimentally, while our bolometric heat equation relies on only four parameters (see supplementary information), two of which (the NbN thermal conductivity and heat-transfer coefficient) can be easily estimated from independent DC electrical measurements. Third, although we over-estimated the cooling of the electrons by the phonons (for assuming the two sub-systems to be in thermal equilibrium), the discrepancy between the reduced temperature of our model and the electron temperature predicted by the 2-$T$ model was partially compensated by under-estimating the cooling of the phonons by the substrate (see supplementary information).

Our model disregards the mechanism of formation and expansion of the photon-induced hotspot [11], so the absorption of a photon results in the immediate superconducting-to-normal transition of a nanowire slab at the center of the nanowire. We assumed this initial normal slab to be as long as the NbN coherence length at zero temperature ($\xi$) and at a temperature (the normal-slab temperature, $T_n$) higher than the substrate temperature ($T_{sub}$). This last assumption was motivated by the fact that if we simulated the photoresponse of an SNSPD at low bias currents ($I_B < 0.7\ I_C$) and at too small a $T_n$ value (for example, $T_n = T_{sub}$), the initial normal slab did not expand and no current was diverted to the load, which was in contrast with the experimental data.

We modeled the thermal coupling between NbN and Sapphire with a state-independent heat-transfer coefficient per unit area with a cubic dependence on temperature $\alpha = A \cdot T^3$ as in ref. [5].

To accurately describe the avalanche formation mechanism in SNAPs, we needed to model the nanowire response to an overcritical current pulse. For this purpose, we inserted a $\xi$-long weak link at the center of the nanowire where the normal domain could nucleate. The weak link was given a slightly lower critical current than the rest of the nanowire ($I_{WL} = 0.999\ I_C$).

When the current through the nanowire exceeds $I_{WL}$, the weak link switches to the normal state (we disregarded the superconducting energy gap suppression time [12]) and is set to a temperature $T_n$.

We present and discuss the parameters used for the electro-thermal simulations (see supplementary information) in the following. The substrate temperature used in the model was $T_{sub} = 4.7$ K, which was based on the temperature measured by a Si diode sensor glued with cryogenic varnish to a detector chip mounted on our cryogenic-device-measurement setup [2]. The nanowires simulated in this paper were 30 nm wide and 4.5 nm thick, like the devices in ref. [2]. The nanowire critical temperature was $T_C = 10.8$ K, which was measured on bare NbN films. The nanowire critical current was $I_C = 7.2$ μA, as we measured on 30-nm-wide-nanowire SNSPDs [2]. The coherence length of our NbN films was assumed to be $\xi = 7$ nm following ref. [6], which reports on similar films. The nanowire inductance per square was $L_\square = 80$ pH/□, which was estimated from the fall time of the photoresponse pulse of 30-nm-wide-nanowire SNSPDs. The nanowire resistance per square was $R_\square = 680$ Ω/□, which was estimated from the ohmic branch of the $I$ - $V$ curves of 30-nm-wide-nanowire SNSPDs measured at 4.7 K. The temperature of the initial normal slab was assumed to be $T_n = 8.5$ K, based on the following criterion: we performed preliminary simulations of the photoresponse of an SNSPD (of inductance $L = 36$ nH) biased at $I_B = 0.6\, I_C$ for a variety of values of $T_n$ (such that $I_C(T_n) < I_B$) and we used the minimum value of $T_n$ for which more than 50% of the bias current was diverted to the load in later device simulations.

As the outcome of our electro-thermal simulations (e.g. the value of $I_{AV}$) was strongly dependent on the strength of the thermal coupling between the NbN film and the sapphire substrate, we estimated the temperature coefficient ($A$) of α from experimental data with the following method: we performed several electro-thermal simulations of a nanowire voltage-biased in *hotspot-plateau* regime [13] (Figure 2.a) varying the value of $A$ to reproduce the behavior observed experimentally.

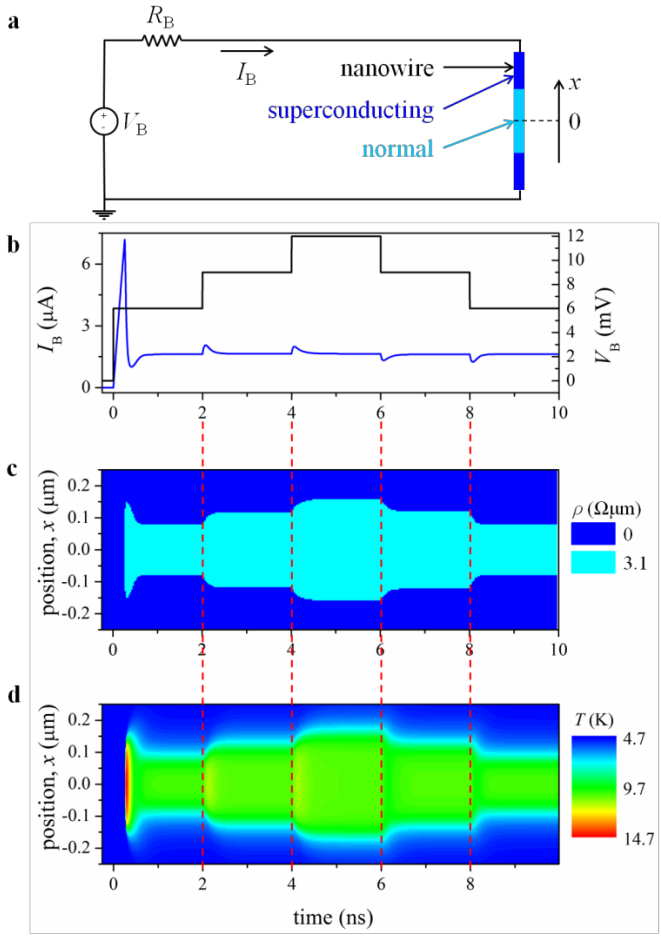

Figure 2. a. Schematics of the electrical circuit used to simulate the hotspot-plateau regime of a 30-nm-wide NbN nanowire. $x$ was the distance from the center of the nanowire. The bias resistor was $R_B = 100\ \Omega$. b. Simulated time evolution of the current ($I_B$, blue curve) and voltage ($V_B$, black curve) in the circuit of Figure a. Red dashed lines mark the instants at which $V_B$ was changed. c, d. Simulated time evolution of the resistivity (c) and temperature (d) profiles along the nanowire.

The circuit simulations started with the nanowire in superconducting state, the bias voltage $V_B = 0$ V, and the current $I_B = 0$ A. At time $= 0$ s, we suddenly increased the bias voltage and as a result, $I_B$ increased until it exceeded the nanowire $I_C$ (Figure 2.b). A normal domain then formed at the center of the nanowire (around the weak link), whose size varied in time (see Figure 2.c) until it stabilized when $I_B$ reached a constant value (the *hotspot current*, $I_{HS}$), as expected for the hotspot-plateau regime [13]. The nanowire responded to any variation in $V_B$ by changing the size of the normal domain and keeping $I_B$ constant and equal to $I_{HS}$, which confirmed that our model correctly describes the hotspot-plateau regime. To find the correct value of $A$ for our nanowires we relied on the fact that the value of $I_{HS}$ depends on the thermal coupling between the NbN film and the sapphire substrate [13]. We used $A$ as a parameter in the hotspot-plateau regime simulations to reproduce a value of $I_{HS}$ matching the hot-spot current measured on 30-nm-wide-nanowire SNSPDs ($I_{HS} = 1.6 \pm 0.1$ μA, extracted from the current-voltage curves measured on 20 devices). We note that we could not use the analytical expression of $I_{HS}$ as a function of $\alpha$ reported in ref. [13] to extrapolate $A$ from the experimental value of $I_{HS}$. Indeed, the expression in ref. [13] was derived under the assumption that given a superconducting nanowire with a stationary spatial temperature profile $T(x)$, a normal domain could exist only where $T(x) > T_C$. We argue that a stationary normal domain could exist under the more general condition that $I_C[T(x)] < I_{HS}$. Indeed, our simulations showed that the stationary normal domain of Figure 2.c was associated with a temperature profile (Figure 2.d) $T(x) < T_C$ everywhere.

To illustrate the capabilities of our model, we simulated the photoresponse of a 2-SNAP. We recorded the time evolution of the temperature and resistivity along the initiating and secondary sections of the device (Figure 3). We also extracted the total resistance of each section and the current through the different parts of the circuit (Figure 4.).

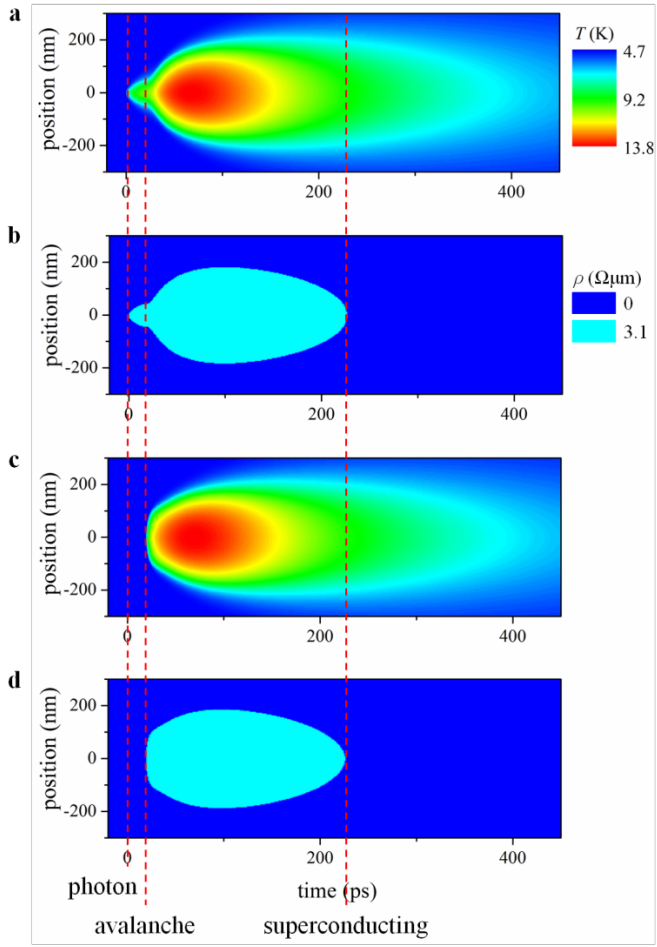

Figure 3. Simulated time evolution of the temperature and resistivity along the initiating (a, b) and secondary (c, d) sections of a 2-SNAP. The device bias current was $I_B = 0.73\ I_C$ The nanowire in inductance was $L_0 = 13.5$ nH. The series inductance was $L_S = 135$ nH. The simulation time step size was 0.1 ps.

To simulate a photon being absorbed in the initiating section (at time = 1 ps), a $\xi$-long slab switches to the normal state (Figure 3.a and b). The normal domain expands due to Joule heating, so the resistance in the initiating section increases (Figure 4.a) and the current through it starts redistributing to the secondary section (Figure 4.b). When the current through the secondary section becomes overcritical (at time = 20 ps) both the initiating and the secondary sections become current-dependent resistors connected in parallel, so their resistance and current fluctuate until they equilibrate, reaching both the same value in all the sections. At this point, the current through the device is redistributed to the read-out (Figure 4.b), until the two sections switch back to the superconducting state (at time = 227 ps).

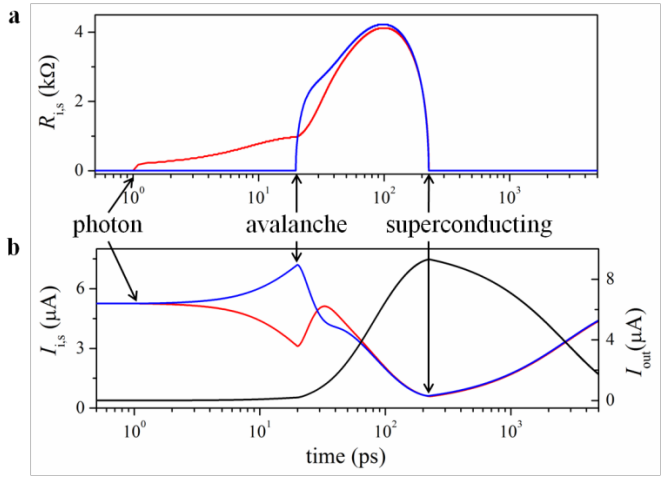

Figure 4. a. Simulated time evolution of the resistance of the initiating ($R_i$, $I_i$, in red) and secondary ($R_s$, $I_s$, in blue) sections of a 2-SNAP. b. Simulated time evolution of the current through the initiating section ($I_i$, in red), the secondary section ($I_s$, in blue) and $R_{load}$ ($I_{out}$, in black).

Performing simulations at different values of the bias current, we could estimate the avalanche current of SNAPs with any number of parallel sections. The values of $I_{AV}$ obtained with our simulations for 2-, 3- and 4-SNAPs were in close quantitative agreement with the experimental values (within the experimental error, see Table 1) [2].

The quality of the agreement between experiment and theory was perhaps surprising given the assumptions in the model (e.g. the fact that we lumped the electron and phonon temperatures in the metal together). This high-quality agreement may be partially coincidental, or may suggest that relative dynamics between the electrons and the phonons does not play a significant role in the device operation.

Table 1. Experimental and simulated values of $I_{AV} / I_{SW}$, where $I_{SW}$ (switching current) is the bias current at which the device switches from the superconducting to the normal state.

|  | 2-SNAP | 3-SNAP | 4-SNAP |
|---|---|---|---|
| $I_{AV} / I_{SW}$ (experiment) | 0.68 ± 0.015 | 0.78 ± 0.02 | 0.84 ± 0.03 |
| $I_{AV} / I_{SW}$ (simulation) | 0.67 | 0.78 | 0.82 |

In conclusion, we developed the electro-thermal model of *N*-SNAPs. Our simulations predicted avalanche currents in agreement with the experimental values and clarified the operation mechanism of these devices.

This work was supported by the Center for Excitonics, under Award Number DE-SC0001088. This work was completed while Prof. K. K. Berggren was on sabbatical at Delft University of Technology, and supported by the Netherlands Organization for Scientific Research. The authors would like to thank Dr A. Annunziata, Dr. E. Driessen, Prof. T. Kalpwijk and Prof. D. Prober for useful scientific discussions.

# Supplementary Information

Francesco Marsili[1,*], Faraz Najafi[1], Charles Herder[1], Karl K. Berggren[1,2]

[1]*Department of Electrical Engineering and Computer Science, Massachusetts Institute of Technology, 77 Massachusetts Avenue, Cambridge, Massachusetts 02139, USA.*

[2]*Kavli Institute of Nanoscience, Delft University of Technology, Lorentzweg 1, 2628CJ Delft, The Netherlands.*

*A. Derivation of the bolometric heat equation*

The general form of the one-dimensional two-temperature heat equations for NbN is [10]:

$$\frac{\partial C_e(T_e)T_e}{\partial t} = -\frac{C_e(T_e)}{\tau_{e\text{-}p}(T_e)}(T_e - T_p) + j^2\rho + \frac{\partial}{\partial x}\left[\kappa_e(T_e)\frac{\partial T_e}{\partial x}\right] \quad (1)$$

$$\frac{\partial C_p(T_p)T_p}{\partial t} = \frac{C_e(T_e)}{\tau_{e\text{-}p}(T_e)}(T_e - T_p) - \frac{C_p(T_p)}{\tau_{esc}}(T_p - T_{sub}) + \frac{\partial}{\partial x}\left[\kappa_p(T_p)\frac{\partial T_p}{\partial x}\right] \quad (2)$$

where $T_e$ and $T_p$ are the electron and phonon temperatures; $C_e$ is the electron specific heat: $C_e \propto T_e$ in the normal state and $C_e \propto \exp[-\Delta(T_e)/k_B T_e]$ in the superconducting state [6]; $C_p \propto T_p^3$ is the phonon specific heat [6]; $\tau_{e-p} \propto T_e^{-1.6}$ is the electron-phonon interaction time [14]; $\tau_{esc}$ is the phonon escape time to the substrate; $\kappa_e$ and $\kappa_p$ are the temperature-dependent electron and phonon thermal conductivities; $\rho$ is the NbN resistivity; $j$ is the nanowire current density.

The bolometric heat equation is obtained adding equation (1) to equation (2) and setting $T_e = T_p = T_r$:

$$\frac{\partial}{\partial t}[C(T_r)T_r] = j^2\rho - \frac{C_p(T_r)}{\tau_{esc}}(T_r - T_{sub}) + \frac{\partial}{\partial x}\left[\kappa(T_r)\frac{\partial T_r}{\partial x}\right] \quad (3)$$

where $T_r$ is the reduced temeperature, $C = C_e + C_p$ and $\kappa = \kappa_e + \kappa_p$.

We further simplified equation (3) into:

$$\frac{\partial}{\partial t}[C(T_r)T_r] = j^2\rho - \frac{\alpha}{d}(T_r - T_{sub}) + \kappa_e(T_r)\frac{\partial^2 T_r}{\partial x^2} \quad (4)$$

neglecting the phonon thermal conductivity and the spatial dependence of $\kappa_e$ ($\partial \kappa_e / \partial x \cdot \partial T_r / \partial x \sim 0$) as in ref. [15]. We note that we expressed the phonon-substrate coupling term in equation (3) in terms of the heat-transfer coefficient per unit area $\alpha = A \cdot T_r^3$ and the film thickness $d$. For the value and temperature dependence of $C$ we followed ref. [6] and we estimated $\kappa_e$ from the nanowire resistivity as in [5].

*B. Comparison between the bolometric and the 2-T heat equations*

In this section we evaluate the validity of our choice of using the bolometric model over the 2-$T$ model to describe the thermal response of our nanowires.

The value of $A / d$ that we estimated from the experimental value of the hotspot current was of the same order of magnitude, but lower than the ratio between the temperature coefficient of $C_p$ ($C_{p0}$) and $\tau_{esc}$ reported in ref. [6] for similar films: $A / d = 67$ W/(mm$^3$K$^4$) ~ $C_{p0} / \tau_{esc} = 125$ W/(mm$^3$K$^4$). The fact that the phonon escape time estimated from our value of $A / d$ (146 ps) was a factor ~ 2 larger than in ref. [6] (78 ps) implied that we under-estimated the cooling of the phonons by the substrate, which in fact partially compensated for over-estimating the cooling of the electrons by the phonons (for assuming thermal equilibrium between the two sub-systems).

To quantitatively support this last claim, we compared the thermal response of a superconducting nanowire to an optical excitation pulse simulated with the bolometric equation and our value of $A$, with the result of the 2-$T$ equations relying on the parameters reported in [6]. For simplicity, we assumed the temperature to be homogeneous along the nanowire and then neglected the thermal conduction terms. The excitation pulse had a peak optical power density of 1.5 mW / µm$^3$ and a duration of 300 ps, which reproduced the joule heating produced by a current of 3 µA flowing through a photon-induced normal domain. The time evolution of the reduced temperature ($T_r$) simulated with the bolometric model was in agreement with the average temperature ($T_{avg}$) between the electron and phonon temperatures simulated with the 2-$T$ model. Considering that the results of the 2-$T$ model were obtained with no free parameters, we concluded that our bolometric model described the nanowire thermal response with an acceptable approximation respect to the more complete model.

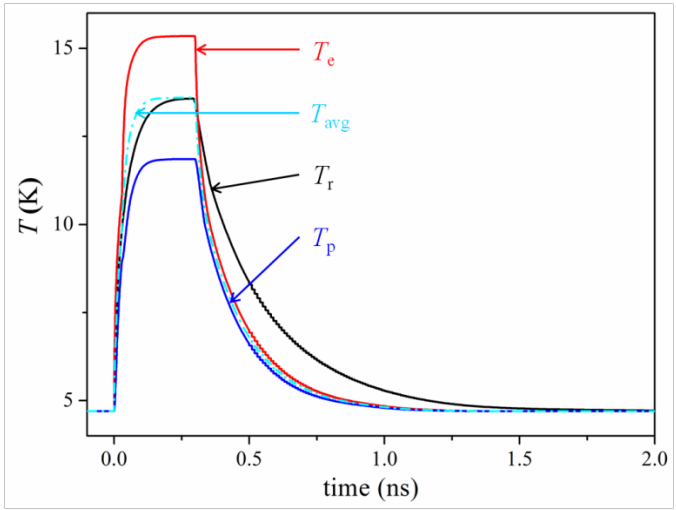

Figure S - 1. Simulated time evolution of the reduced temperature ($T_r$, in black), the electron temperature ($T_e$, in red), the phonon temperature ($T_p$, in blue), and the average between $T_e$ and $T_p$ ($T_{avg}$, in blue).

## C. Parameters of the electro-thermal simulation.

Table SI - 1. Parameters of the electro-thermal simulation.

| Symbol | Quantity | Value |
|---|---|---|
| $T_{sub}$ | substrate temperature | 4.7 K |
| $T_C$ | critical temperature | 10.8 K |
| $I_C$ | critical current | 7.2 µA |
| $\xi$ | Ginzburg-Landau coherence length | 7 nm |
| $L_\square$ | kinetic inductance per square | 80 pH/□ |
| $R_\square$ | resistance per square | 680 Ω/□ |
| $T_n$ | normal-slab temperature | 8.5 K |
| $A$ | Temperature coefficient of the heat-transfer coefficient per unit area | 300 W/(m²K⁴) |